\documentclass[12pt,a4]{article}

\newcommand{\beq}{\begin{equation}}
\newcommand{\eeq}{\end{equation}}
\newcommand{\beqa}{\begin{eqnarray}}
\newcommand{\eeqa}{\end{eqnarray}}
\newcommand{\nn}{\nonumber}

\begin{document}
\begin{titlepage}
\rightline{DSF-T-45/99}
\rightline{NORDITA 1999/78-HE}
\rightline{\hfill December 1999}
 
\vskip 1.2cm
 
\centerline{\Large \bf Two-Loop $\Phi^{4}$-Diagrams from String Theory}  
 
\vskip 1.2cm

\centerline{\bf Raffaele Marotta}
\centerline{\sl NORDITA}
\centerline{\sl Blegdamsvej 17, DK-2100 Copenhagen \O, Denmark}
\centerline{\sl marotta@nbi.dk}
 
\vskip .2cm
 
\centerline{\bf Franco Pezzella}
\centerline{\sl I.N.F.N., Sezione di Napoli}
\centerline{\sl and Dipartimento di Scienze Fisiche, Universit\`a di Napoli}
\centerline{\sl Complesso Universitario di Monte S. Angelo, Ed. G, I-80126 Napoli, Italy}
\centerline{\sl pezzella@na.infn.it}

\vskip 2cm

\begin{abstract}

Using the {\em cutting and sewing} procedure we show how to get 
Feynman diagrams, up to two-loop order, of $\Phi^{4}$-theory
with an internal $SU(N)$ symmetry group, starting from 
tachyon amplitudes of the open bosonic string theory. In a properly defined 
field theory limit, we easily identify the corners of the string moduli space 
reproducing the correctly normalized field theory amplitudes expressed
in the Schwinger  parametrization.
\footnotetext[1]{This research was partially supported by the EU-TMR programme 
ERBFMRX-CT96-0045 (Nordita-Copenhagen and Universit\`a di Napoli)}

\end{abstract}

\end{titlepage}

\newpage

\section{Introduction}

Much interest has been devoted to the relation between string theory and
field theory over the last years. It is well-known that any
string theory reduces to an effective field theory in the so-called
{\em zero-slope limit} where
the inverse string tension $\alpha' \rightarrow 0$. The latter is a
physical dimensional parameter acting as an ultraviolet cutoff in the
integrals over loop momenta and so makes multiloop amplitudes free from 
ultraviolet divergences. This is a basic reason why a string theory 
can provide a 
consistent quantum theory of gravity, unified with non-abelian gauge theories.

Furthermore, string theory also manages to organize scattering amplitudes
in a very compact form, which makes much easier to calculate non-abelian
gauge theory amplitudes by starting from a string theory and performing
the zero-slope limit, rather than using traditional field theory techniques.
We would like here to add that the expression of string amplitudes is known
explicitly, including also the measure of integration on 
moduli space, in the case of the bosonic open string for an arbitrary
perturbative order \cite{D}. 

These are the features of string theory that have led some authors
to use it as an efficient tool to compute gluon amplitudes in Yang-Mills
theory \cite{BK1} $\div$ \cite{DMLRM} or in quantum gravity, 
where the improvement over traditional 
techniques was even more spectacular \cite{Bgr}, and light has been shed 
on the perturbative relations between gravity and gauge theory \cite{BDDPR}. 
We would like to add that these string-methods have also inspired
some authors in developing very interesting techniques based on the world-line
path-integrals \cite{kajsch}. 

The general procedure on which the derivation of field theory amplitudes
from string amplitudes is
based consists in starting, for example in the case of Yang-Mills amplitudes, 
from a given multiloop gluon string amplitude 
and in singling out different regions of the moduli space that, in the 
low-energy limit, reproduce different field theory diagrams. This
program has been carried out at one-loop \cite{BDK1} \cite{DMLMR} and, 
at this order, the five-gluon amplitude has been obtained for the 
first time \cite{BDK2}.

Some attempts have been performed for extending to the two-loop case
this procedure \cite{DMLMR} \cite{MR} \cite{BRY} which while, on the one hand,
does not present many difficulties in computing
Yang-Mills vacuum diagrams, on the other hand becomes difficult 
to handle when treating amplitudes with external states, because of their complicated structure.
In order to avoid the computational difficulties associated with this kind of 
multiloop   Yang-Mills amplitudes, which are inessential for the understanding
of the field theory limit, one can consider amplitudes involving scalar 
particles. In fact, differently from what happens in gauge theories, in string
theory
there is not a big conceptual difference between gluon and scalar diagrams.
Therefore one can get from scalar amplitudes, with much less computational
cost, the whole information about the
corners of the moduli space reproducing the known field theoretical results:
these regions are exactly the ones giving the correct field theory diagrams
also in the case of gluon amplitudes.

The scalar particles one is referring are, of course,  
the tachyons of the bosonic string theory. So one can  consider 
a slightly different zero-slope limit of the bosonic
string in which only the lowest tachyonic state, with a mass satisfying
$m^{2}= -1/\alpha'$ is kept. In the case of tree and one-loop diagrams,
this procedure is equivalent to take the zero-slope limit of an old
pre-string dual model characterized by an arbitrary value of the intercept
of the Regge trajectory. It was recognized the inconsistency of this model,
but the field theory limit of tree and one-loop diagrams of this pre-string
dual model was shown to lead to the Feynman diagrams of $\Phi^{3}$ 
theory \cite{S}.

In a previous paper \cite{DMLMR}, it has been explicitly shown that
by performing the zero-slope limit as above explained,
one correctly reproduces the Feynman diagrams of $\Phi^{3}$ theory, up
to two-loop order.
In this way it has been provided an algorithm generalizable to
Yang-Mills theory in order to obtain all diagrams containing the three-gluon 
vertices.  

This paper is the natural development of the program carried out in Ref. 
\cite{DMLMR}, its aim being to extend the conceptual scheme 
pursued in it, to two-loop amplitudes containing four-gluon vertices, 
whose very complicated
structure, when external particles are involved, has not 
allowed to get meaningful results up to now. This means that, exploiting
the above mentioned analogy between scalar and gluon amplitudes, one could
start from string amplitudes involving tachyons, perform on them the 
field theory limit in which their mass is fixed and correctly identify 
the corners associated to the different field theory diagrams
of $\Phi^{4}$ theory. As observed in Ref. \cite{DMLMR}, this
analysis may be a bit lengthy, since there are many corners of the string
moduli space that contribute to the same field theory diagrams. 
But one can pursue an alternative and equivalent method, the {\em
sewing and cutting} procedure, that also leads to the correct identification
of field theory diagrams and that has been successfully applied to 
get two-loop $\Phi^{3}$ amplitudes in \cite{DMLMR}. This is indeed the
technique we use in this paper. 
We would like here to observe that, until now, four-vertices have been
obtained as finite remainders of tachyon exchange \cite{MR} \cite{MP}
and that one of the main
peculiarities of our technique is that the identification of the
right corner in the moduli space is independent on any handling
tachyon exchange. Hence it is reasonable to think that the {\em sewing and
cutting} procedure
is easily extendible to consistent string theories.

Finally, another interesting motivation to analyze the field theory limit of
scalar amplitudes is understanding the analogous limit 
in a string theory with a non-zero $B$-field. This limit
yields non-commutative field theories and provides a hope that it would
be possible to cure quantum field theory divergences \cite{SW}
\cite{ABK}.

The paper is organised as follows. 

In sect. 2 we present the operator formalism that we use for computing
multiloop scalar amplitudes in the bosonic open string theory. We consider 
scalar particles having an internal $SU(N)$
symmetry group and perform a large $N$ limit of the relative amplitudes
so that only planar diagrams are taken into account. 
 
In sect. 3 we define the {\em sewing and cutting} procedure applying
it to the four-tachyon tree amplitude. We define a proper field theory
limit where cubic and quartic interactions are reproduced.
 
In sect. 4 we check the validity of this procedure by deriving
from the two- and four-tachyon amplitudes at one-loop, respectively,
the {\em tadpole} and the {\em candy} diagram of $\Phi^4$ theory.

In sect. 5 the {\em sunset} and the {\em double-candy}
diagrams are computed. Each field-theory diagram is obtained with 
its own right overall normalization.

\section{Multiloop scalar amplitudes in the bosonic open string}

The planar $h$-loop scattering amplitude of $M$ tachyons with momenta
$p_{1}, \dots, p_{M}$ is:
\beqa
A_{M}^{(h)}(p_{1}, \dots, p_{M})  =   N^{h} \mbox{Tr} 
\left[\lambda^{a_{1}} \cdots
\lambda^{a_{M}} \right] C_{h} \left[ 2 g_{s} (2 \alpha')^{\frac{d-2}{4}}
\right]^{M} \nonumber \\
 \times \int \left[ dm \right]^{M}_{h} \prod_{i<j} \left[ 
\frac{ \exp{ \left( {\cal G}^{(h)} ( z_{i}, z_{j} ) \right)}}{ 
\sqrt{ V'_i(0)V'_j(0) } } 
\right]^{2 \alpha' p_{i} \cdot p_{j} }
\label{AMh}
\eeqa
where $N^{h} \mbox{Tr} (\lambda^{a_1} \cdots \lambda^{a_M})$ is the
appropriate $SU(N)$ Chan-Paton factor, with the $\lambda$'s being the
generators of $SU(N)$ in the fundamental representation, normalized as
\beq
\mbox{Tr} (\lambda^{a} \lambda^{b}) = \frac{1}{2} \delta^{ab}  \,\,\, ,
\eeq 
$g_s$ is the dimensionless string coupling constant, 
$C_{h}$ is a normalization factor given
by:
\beq
C_{h} = \frac{1}{(2 \pi)^{dh}} g_{s}^{2h-2} \frac{1}{(2 \alpha')^{d/2}}
\eeq
and ${\cal G}^{(h)}$ is the $h$-loop bosonic Green function
\beq
{\cal G}^{(h)} (z_i, z_j) = \mbox{log} E^{(h)} (z_i,z_j) - \frac{1}{2}
\int_{z_i}^{z_j} \omega^{\mu} (2 \pi {\mbox Im} \tau_{\mu \nu} )^{-1}
\int_{z_{i}}^{z_j} \omega^{\nu}   ,
\eeq
with $E^{(h)} (z_{i},z_{j})$ being the prime form, $\omega^{\mu} (\mu=1,
\cdots, h)$ the abelian differentials and $\tau_{\mu \nu}$ the period
matrix. All these geometrical objects, which are peculiar of the open Riemann
surface of genus $h$ on which the amplitude is defined, can be
explicitly written in the Schottky parametrization of the surface itself, and
their expressions can be found for example in Ref. \cite{DPFHLS}. 
The measure $\left[ dm \right]^{M}_{h}$, when written in
terms of the Schottky parameters, is given by:

\beqa
\left[ dm \right]^{M}_h & = & \frac{1}{dV_{abc}} \prod_{i=1}^{M}
\frac{dz_i}{V'_{i}(0)}
\prod_{\mu=1}^{h} \left[ \frac{dk_{\mu} d \xi_{\mu} d \eta_{\mu}}{k^{2}_{\mu}
(\xi_{\mu} - \eta_{\mu} )^{2}} (1-k_{\mu})^{2} \right] \nonumber \\
{} & \times & \left[ \mbox{det} (-i \tau_{\mu \nu} ) \right]^{-d/2}
\prod_{\alpha} \,{'} \left[ \prod_{n=1}^{\infty} (1-k^{n}_{\alpha})^{-d}
\prod_{n=2}^{\infty} (1-k_{\alpha}^{n} )^{2} \right]
\eeqa
where $k_{\mu}$ are the multipliers, $\xi_{\mu}$ and $\eta_{\mu}$
the fixed points of the generators of the Schottky group and $dV_{abc}$
is the projective invariant volume element:
\[
dV_{abc} = \frac{ d \rho_{a} d \rho_{b} d \rho_{c}}{ (\rho_{a} - \rho_{b})
(\rho_{a} - \rho_{c} ) (\rho_{b} - \rho_{c} )} \,\,\, ,
\] 
with $\rho_{a}$, $\rho_{b}$, $\rho_{c}$ being any three of the $M$ 
Koba-Nielsen variables $z_i$, or of the $2h$ fixed points of the 
generators of the Schottky group, which can be fixed at will; finally, 
the primed product over $\alpha$ denotes the product over the primary
classes of elements of the Schottky group \cite{DPFHLS}. Furthermore, like 
in any planar open string amplitude, the Koba-Nielsen variables 
must be cyclically ordered along one of the boundaries of the world-sheet, 
for example according to:
\beq
z_{1} \geq z_{2} \geq \cdots \geq z_{M}
\eeq
and the ordering of Koba-Nielsen variables automatically prescribes the
one of color indices.

The presence in (\ref{AMh}) of terms involving the quantity $V'_{i}(0)$
originates from the $M$-point $h$-loop vertex $V_{M;h}$ of the operator
formalism \cite{D}. This can be regarded as a generating functional for
scattering amplitudes among arbitrary states, at all orders in perturbation
theory. In fact, by saturating the operator $V_{M;h}$ with $M$ states
$|\alpha_{1}>, \dots, |\alpha_{M}>$, one obtains the corresponding
amplitude:
\beq
A^{(h)} (\alpha_{1}, \cdots, \alpha_{M}) = V_{M;h} |\alpha_{1}> \dots
|\alpha_{M}>.
\eeq

The explicit expression of $V_{M;h}$ for planar diagrams of the
open string depends on $M$ real Koba-Nielsen
variables $z_{i}$ through $M$ projective transformations $V_{i}(z)$,
which define local coordinate systems vanishing around each $z_{i}$,
i.e. such that:
\beq
V_{i}(0) = z_{i} \,\, .
\eeq

When $V_{M;h}$ is saturated with $M$ physical string states satisfying
the mass-shell condition, the corresponding amplitude does not depend on
the $V_{i}$'s. Such a dependence remains for off-shell string amplitudes
and for these it is therefore necessary to make a choice of the projective
transformations $V_{i}$'s \cite{DMLRM} \cite{DMLMR}. It has been shown
that a principle guiding such a choice is the requirement of projective
invariance of off-shell string amplitudes \cite{CMPP}.   

Let us denote by
$|p>$ a state representing a tachyon with momentum $p$. It is
created by the vertex operator
\beq
V(z) = {\cal N}_{t} :e^{i \sqrt{2 \alpha'} p \cdot X(z)} :
\eeq
where colons denote the standard normal ordering on the modes of the open 
string coordinate $X(z)$ and ${\cal N}_{t}$ is a normalization factor 
\cite{DMLMR}:
\beq
{\cal N}_{t} = 2 g_{s} ( 2 \alpha')^{\frac{d-2}{4}} .
\eeq  
If we write, as usual 
\beq
X^{\mu}(z) = \hat{q}^{\mu} - i \hat{p}^{\mu} \mbox{log}z + i \sum_{n \neq 0}
\frac{\hat{a}^{\mu}_{n}}{n} z^{-n}
\eeq
then the tachyon state is
\beq
|p> \equiv \lim_{z \rightarrow 0} V(z)|0> = {\cal N}_{t} 
e^{i p \cdot \hat{q}}|0> \,\,\,  .
                       \label{tachyon}
\eeq
The tachyon state is {\em on-shell} if
\beq
      p^{2} = -m^{2} = \frac{1}{\alpha'} \,\,\,  .
\eeq

The amplitude (\ref{AMh}) is obtained by saturating $V_{M;h}$ on $M$
tachyon states defined in (\ref{tachyon}). 

\section{Tree diagrams from the {\em sewing and cutting} technique}

\subsection{Tree $\Phi^{3}$-diagrams}
   
\setcounter{equation}{0}
\indent

\input epsf

\begin{figure}[h]
\begin{center}
\parbox{3.5 cm}{\epsfxsize 3.5cm \epsfbox{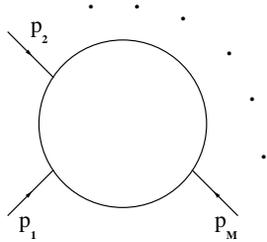}}
\end{center}
\caption{Planar tree scattering amplitude}
\label{tree}
\end{figure}

The planar tree scattering amplitude of $M$ on-shell bosonic open
string tachyons with momenta $p_{1}, \dots, p_{M}$ each satisfying the 
mass-shell condition $p^{2}=-m^{2}=\frac{1}{\alpha'}$ is obtained from
the {\em master equation} in (\ref{AMh}) considered for $h=0$: 
\beq
A_{M}^{(0)}(p_{1}, \dots, p_{M}) = \mbox{Tr} \left[ \lambda^{a_{1}} \cdots
\lambda^{a_{M}} \right] C_{0} {\cal N}_{t}^{M} \int \prod_{i=1}^{M} \frac{dz_{i}} 
{dV_{abc}} \prod_{i < j} (z_{i} - z_{j})^{2 \alpha' p_{i} \cdot p_{j} } .
                                                  \label{AM}
\eeq
The corresponding diagram is depicted in Fig. (1).

In the case of four on-shell tachyons Eq. (\ref{AM}) becomes
\beq
A_{4}^{(0)}(p_{1},p_2,p_3, p_{4}) = \mbox{Tr} \left[ \lambda^{a_{1}} 
\lambda^{a_{2}} \lambda^{a_{3}}
\lambda^{a_{4}} \right] C_{0} {\cal N}_{t}^{4}
\int_{0}^{1} dz (1-z)^{2 \alpha' p_{2} \cdot p_{3}} z^{2 \alpha' p_{3} \cdot 
p_{4}} 
                                                  \label{A4}
\eeq
where $z_{1}$, $z_{2}$ and $z_{4}$ have been respectively fixed at
 $ +\infty,1$ and $0$.

In the limit $\alpha' \rightarrow 0$ the amplitude in Eq. (\ref{A4})
yields the tree level Feynman diagrams of scalar field theories:
different regions in moduli space lead to different field theory diagrams.
In fact, according to the corner of moduli space where the low-energy limit is
performed, one can recover, for instance, $\Phi^{3}$- or $\Phi^{4}$-scalar
diagrams.

In order to understand which regions in moduli space lead
to the different field theory diagrams, one can use the so-called
{\em sewing} {\em and} {\em cutting} procedure. This consists in
starting from a string diagram
and in cutting it in three-point
vertices; next we fix the legs of each three-point vertex at 
$ + \infty$, $1$ and $0$. Then we reconnect 
the diagram by inserting between two three-point vertices a 
suitable propagator acting as a well specified projective transformation.
This is chosen in such a way that its fixed points are just the 
Koba-Nielsen variables of the two legs that have to be sewn. The geometric
role of the propagator is to identify the local coordinate systems defined
around the punctures to be sewn.

\begin{figure}[h]
\begin{center}
\parbox{11 cm}{\epsfxsize 11cm \epsfbox{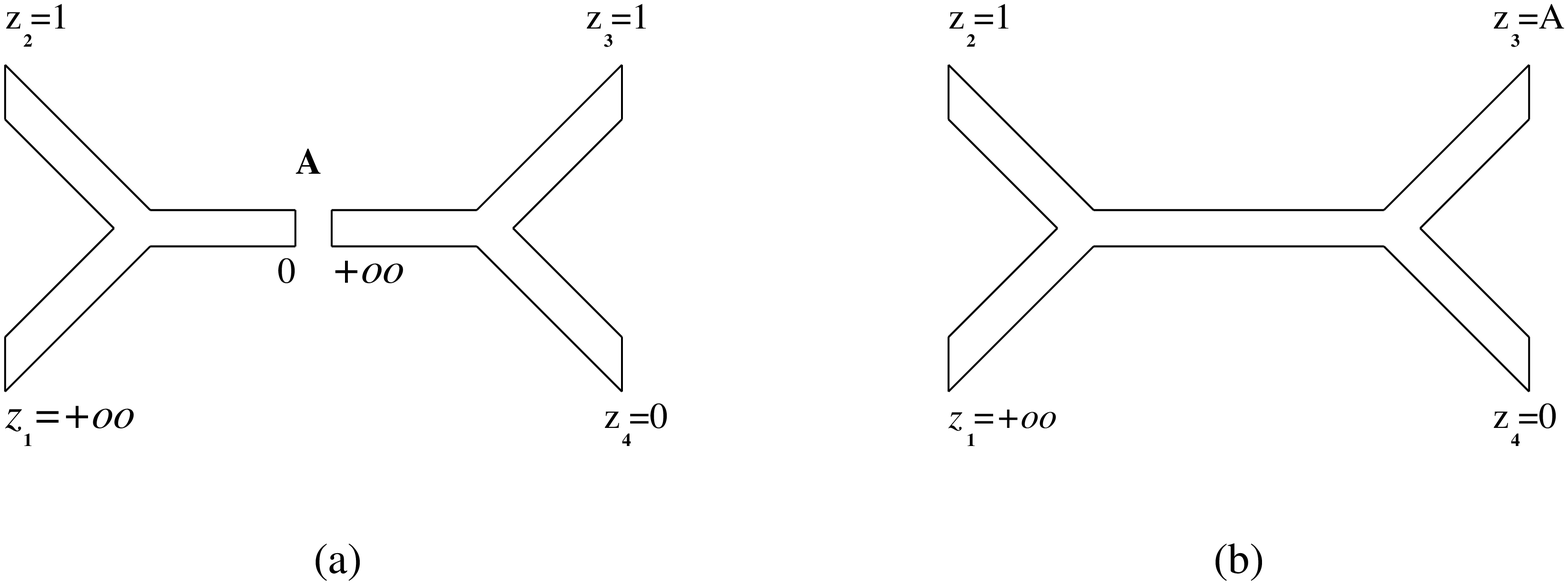}}
\end{center}
\caption{Sewing of two three vertices in the s-channel.}
\label{trees}
\end{figure}

For illustrating in a simple case how this technique works, 
we show how the four-point tree diagrams of $\Phi^{3}$-theory can be generated. 
The starting point is the four-tachyon
tree string diagram. This can be obtained by 
sewing two three-point vertices as shown in Fig. (\ref{trees}). 
We sew the leg corresponding to the point $0$ in the vertex at the left
hand in Fig. (2a) to the leg corresponding to the point $+\infty$ in the one at the
right hand through a propagator corresponding to the projective transformation
\beq
S(z)=Az    \label{prop}
\eeq
which has $0$ and $+\infty$ as fixed points and the parameter $A$, with
$0 \leq A \leq 1$, as multiplier. Performing the sewing means,
in this procedure, to transform {\em only} the punctures of the three-point 
vertex at the right hand in Fig. (2a) through (\ref{prop}), hence the puncture 
$z_{3}=1$ transforms into $S(1)=A$ while the other two punctures remain
unchanged. 

\input epsf

In general, after the sewing has been 
performed, the Koba-Nielsen variables become
functions of the parameter $A$ appearing in the projective transformation
(\ref{prop}). It is possible to give a simple geometric
interpretation to this parameter, if a correspondence is established
between the projective transformation in Eq. (\ref{prop}) and the 
string propagator, written in terms of the operator 
$e^{-\tau (L_{0}-1)}$. The
latter indeed propagates an open string through
imaginary time $\tau$ and creates a strip of length $\tau$. In fact
the change of variable 
$z=e^{-\tau}$ allows the string propagator to be written as
\beqa
\frac{1}{L_{0}-1} &  = & \int_{0}^{1} dz   z^{L_{0} - 2} \nonumber \\
{} & = &  \int_{0}^{\infty} d \tau \exp \left( -\tau \alpha' \left[ p^{2} + 
\frac{1}{\alpha'} (N-1) \right] \right) \label{LO-1}
\eeqa
and to establish the following relation between $\tau$ and $A$: 
\beq
\tau = - \mbox{log}A .   \label{logA}
\eeq  

The multiplier $A$ results to be therefore related to the length of the strip
connecting two three-vertices.

On the other hand, since we want to reproduce $\Phi^{3}$-theory diagrams
we have to consider a low-energy limit of string amplitudes in which only
tachyons propagate as intermediate states. This is achieved observing
from (\ref{LO-1}) that the only surviving contribution in the limit
$\alpha' \rightarrow 0$ with $\tau \alpha'$ kept fixed is the one coming
from the level $N=0$, i.e. from tachyons with fixed mass given by
$m^{2}= -\frac{1}{\alpha'}$. It is obvious that this also corresponds
to the limit $\tau \rightarrow \infty$ and hence, from (\ref{logA}),
to $ A \rightarrow 0$. From these considerations it seems
natural to introduce the variable $x= \tau \alpha'$ in terms of which
the string propagator (\ref{LO-1}), reproduces, in the above mentioned
limit, the scalar propagator
\[
\frac{1}{p^{2}+m^{2}} = \int_{0}^{\infty} dx e^{-x(p^{2} + m^{2})}
\]
with $x$ being interpreted as the Schwinger proper time.

From a geometrical point of view, one can imagine that the
the strip connecting the two three-vertices, in this field theory limit, 
becomes ``very long and thin'', so that only the lightest states propagate.

Let us now rewrite the amplitude (\ref{A4}) in terms of the Schwinger parameter
$x$ or, equivalently, in terms of the multiplier $A$:
\beqa
A_{4}^{(0)} & = & \mbox{Tr} \left[ \lambda^{a_{1}} 
\lambda^{a_{2}} \lambda^{a_{3}} \lambda^{a_{4}} \right]
\frac{{\cal N}_{t}^{4}}{g_{s}^{2} (2 \alpha^{'} )^{d/2}} \int_{0}^{\epsilon}
dA \exp {(2\alpha^{'} p_{3}p_{4} \mbox{log} A) }  \nonumber \\
{} & = & \frac{1}{8} \mbox{Tr} \left[ \lambda^{a_{1}} \lambda^{a_{2}}
\lambda^{a_{3}} \lambda^{a_{4}} 
\right]
\frac{g^{2}_{\phi^{3}}}{\left[ (p_{1} + p_{2})^{2} + m^{2} \right] } 
                                         \label{A43}
\eeqa
where it has been used the well-known relation between $g_{s}$ and 
$g_{\phi^{3}}$ \cite{DMLMR}:
\beq
g_{\phi^{3}} = 16 g_{s} (2 \alpha^{'} )^{ \frac{d-6}{4} } .
\eeq

By performing the sum over inequivalent permutations and selecting
those diagrams which contribute to the $s$-channel, we get

\beq
A^{(0){s-channel}}_{4}(p_{1},p_{2},p_{3}, p_4)  = 
 \frac{1}{8} G
\frac{g^{2}_{\phi^{3}}}{\left[ (p_{1} + p_{2})^{2} + m^{2} \right] } \label{B}
\eeq
where
\beq
G \equiv \mbox{Tr} \left[ \lambda^{a_{1}} \lambda^{a_{2}} \lambda^{a_{3}} 
\lambda^{a_{4}} \right] + \mbox{Tr} \left[ \lambda^{a_{1}} \lambda^{a_{2}} 
\lambda^{a_{4}} 
\lambda^{a_{3}} \right] + 
\mbox{Tr} \left[ \lambda^{a_{1}} \lambda^{a_{3}} \lambda^{a_{4}} 
\lambda^{a_{2}} \right] + \mbox{Tr} \left[ \lambda^{a_{1}} \lambda^{a_{4}} \lambda^{a_{3}} 
\lambda^{a_{2}} \right]  .
\eeq
Taking into account the relation satisfied by the $SU(N)$ generators 
$\lambda^{a}$ $(a=1, \dots, N^{2}-1)$ in the $N$-dimensional fundamental
representation:
\[
\left\{ \lambda^{a}, \lambda^{b} \right\} = \frac{1}{N} \delta_{ab} + 
d^{abc} \lambda^{c}
\]
and the one satisfied by the totally symmetric tensor
$d^{abc}$
\[
d^{abc}= 2 \mbox{Tr} \left[ \left\{ \lambda^{a}, \lambda^{b} \right\} 
\lambda^{c} \right], 
\]
one can get, in the large $N$ limit
\beq
G = \frac{1}{2} d^{a_{1} a_{2} l} d^{a_{3} a_{4} l}
\eeq
where a sum on $l$ is understood.
Hence
\beq
A^{(0)s-channel}_{4}  (p_{1},p_{2},p_{3}, p_{4})  =  \frac{1}{16} 
\frac{g^{2}_{\phi^{3}}}{\left[ (p_{1} + p_{2})^{2} + m^{2} \right] }
d^{a_{1} a_{2} l} d^{a_{3} a_{4} l} . \label{A40}
\eeq

The result (\ref{A40}) can be also obtained by using standard techniques.
Indeed, taking into account the equation
\beq
\int_{0}^{1} (1-z)^{ 2 \alpha' p_{2} \cdot p_{3} } z^{ 2 \alpha' p_{3} \cdot
p_{4} }  =  B \left[ 2 \alpha' p_{2} \cdot p_{3} + 1, 2 \alpha' p_{3} 
\cdot p_{4} + 1 \right]
\eeq
with $B$ being the Euler Beta function, 
the following amplitude is obtained in the limit $\alpha' \rightarrow 0$:
\beq
A_{4}^{(0)} (p_{1},p_{2},p_{3}, p_{4}) = \mbox{Tr} \left[ \lambda^{a_{1}}
\lambda^{a_{2}} \lambda^{a_{3}} \lambda^{a_{4}} \right] C_{0} {\cal N}_{t}^{4}
\frac{1}{\alpha'} \left[ \frac{1}{-s + m^{2}} + \frac{1}{-t+m^{2}}
\right]
\eeq

\begin{figure}[h]
\begin{center}
\leavevmode
\parbox{9 cm}{\epsfxsize 9cm \epsfbox{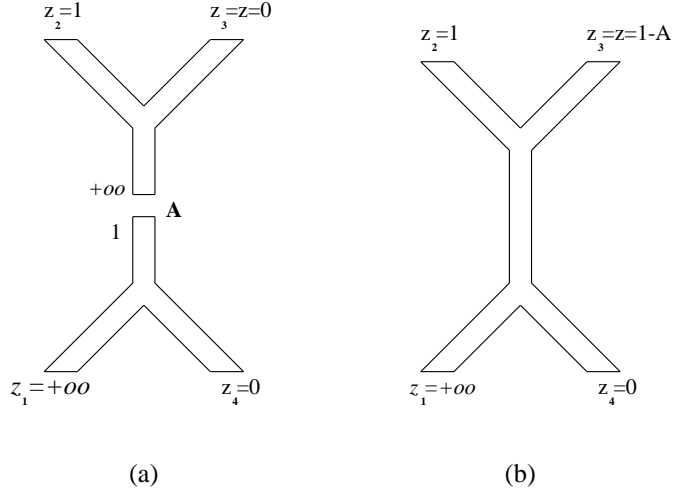}}
\end{center}
\caption{Sewing of two three vertices in the t-channel}
\label{treet}
\end{figure}

The first term on the right hand reproduces the same result as in 
Eq. (\ref{A40}), including the overall factor. The second term, giving
the amplitude in the $t$-channel, can be also obtained by the 
sewing and cutting procedure as it is shown in Fig. (\ref{treet}).

\input epsf

In this case one has to sew the leg corresponding to $+\infty$ in the upper
vertex in Fig. (3a) to the leg corresponding to $1$ in the lower vertex
through the transformation
\[
S(z)=Az+1-A
\]
having $1$ and $+\infty$ as fixed points and which transforms the puncture
$z_{3}=0$ of the upper vertex into $S(0)=1-A$.
Therefore the amplitude (\ref{A4}), in the limit $A \rightarrow 0$, becomes:
\beqa
A_{4}^{(0)}(p_{1}, \dots, p_{4}) & = & \mbox{Tr} \left[ \lambda^{a_{1}} \cdots
\lambda^{a_{4}} \right] C_{0} {\cal N}_{t}^{4}
\int_{0}^{\epsilon} dA 
\exp{\left( 2 \alpha' p_{2} \cdot p_{3} \mbox{log} A \right)} \nonumber\\
{} & = & C_{0} {\cal N}_{t}^{4}
       \frac{1}{\alpha'} \left[ \frac{1}{-t+m^{2}}\right]                                           \label{A}
\eeqa

It is straightforward to show that the sum of the amplitudes 
(\ref{B}) and (\ref{A}) coincides with the 
four-point Green function in the field theory defined by the following
action:
\beq
S  =  \int d^{d} x \left[ \frac{1}{2} \partial_{\mu} \phi^{a} \partial^{\mu} \phi^{a} + \frac{1}{2}
m^{2} \phi^{a} \phi^{a} - \frac{d^{abc}}{4} \frac{g_{\phi^{3}}}{3!} \phi^{a} \phi^{b}
\phi^{c} \right] 
\eeq

\subsection{Tree scalar $\Phi^{4}$-diagrams from String Amplitudes}
The starting point is also in this case the amplitude (\ref{AM}) that, from
(\ref{AMh}), can be expressed
also in terms of the Green functions ${\cal G}^{(0)}(z_{i}, z_{j})$,
defined on the world-sheet in the following way:
\beq
{\cal G}  ^{(0)}(z_{i},z_{j}) = \mbox{log} \left( z_{i} - z_{j} \right)
                               \label{Gf}
\eeq

Our aim is now to consider a suitable limit of the string four-tachyon
amplitude which can reproduce the diagram
corresponding to the tree four-point vertex of $\Phi^{4}$-theory.
With reference again to the Fig. (\ref{trees}), this diagram has 
to correspond to a limit in which the length of the tube
connecting the two three-vertices composing the string diagram goes to
zero in the limit $\alpha' \rightarrow 0$, i.e.
\[
\tau = \frac{x}{\alpha'}= - \mbox{log} A \rightarrow 0
\]
This corresponds to the limit $A \rightarrow 1$, and hence $z \rightarrow 1$
or, equivalently, $z_{3} \rightarrow z_{2}$.

In this limit the Green function ${\cal G}^{(0)}(z_{2},z_{3})$ is divergent.
We regularize it by introducing a  cut-off $\epsilon$
on the world-sheet so that
\[
\lim_{z_{2} \rightarrow z_{3}} \mbox{log} \left[ (z_{2}-z_{3})+ \epsilon
\right]  = \mbox{log}
\epsilon
\]
and
\[
\lim_{\alpha' \rightarrow 0} \alpha' \mbox{log} \epsilon = 0
\]
We consider therefore the amplitude $A_{4}^{(0)}$ in Eq. 
(\ref{A4}) in the field theory limit
defined by: 
\[
A=z=1-\epsilon,
\]
\beqa
\alpha' \rightarrow 0 & \mbox{and} & x =  -\alpha' \mbox{log} 
\epsilon \rightarrow 0.   \label{lim}
\eeqa
    
in which it reduces to
\beqa
A_{4}^{(0)} (p_1, \dots, p_4) & = & \mbox{Tr} \left[ \lambda^{a_{1}} \lambda^{a_{2}} 
\lambda^{a_{3}} \lambda^{a_{4}} \right] C_{0} {\cal N}_{t}^{4}
\int_{0}^{1} dz e^{ 2 \alpha' p_{2} \cdot p_{3}  \log \epsilon}
\nonumber \\
{} & =  & 2^{4} \mbox{Tr} \left[ \lambda^{a_{1}} 
\lambda^{a_{2}} \lambda^{a_{3}}
\lambda^{a_{4}} \right] g_{s}^{2} (2 \alpha')^{\frac{d-4}{2}} \label{A44}.
\eeqa
The complete amplitude is obtained by performing the sum over 
non cyclic permutations, finally getting
\beq
A_{4}^{(0) complete}   =  4 g^{2}_{s} (2 \alpha')^{\frac{d-4}{2}}
 \left\{ d^{ a_{1} a_{2} a_{c}} d^{a_{3} a_{4} a_{c}}
+ d^{a_{1}a_{3}a_{c}}d^{a_{c}a_{2}a_{4}} + d^{a_{1}a_{4}a_{c}}d^{a_{c}a_{2}
a_{3}} \right\} . \label{A4c}
\eeq
where a sum on repeated indices is understood.

We compare the result in (\ref{A4c}) with the color ordered vertex generated
by the following scalar field theory:
\beq
{\cal L} = \mbox{Tr} \left[ \partial^{\mu} \phi \partial_{\mu} \phi + m^{2} \phi^2
- \frac{g_{\phi^4}}{4!} \phi^4 \right] \label{phi4}
\eeq
obtaining the {\em matching condition}
\beq
g_{\phi^4} = 4 g_{s} (2 \alpha')^{d/2 - 2} . \label{matching}
\eeq

Eqs. (\ref{A43}) and 
(\ref{A44}) show that amplitudes of $\Phi^{3}$-theory are obtained at the
boundaries of the integration region, while the ones of $\Phi^{4}$-theory
get contributions from the whole integration region.

\section{One-loop $\Phi^{4}$-diagrams from string amplitudes}
\setcounter{equation}{0}
\subsection{Tadople diagram}

\input epsf

\begin{figure}[h]
\begin{center}
\parbox{3 cm}{\epsfxsize 3cm \epsfbox{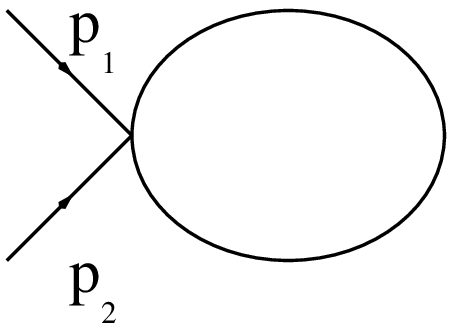}}
\end{center}
\caption{Tadpole}
\label{tad}
\end{figure}

In this subsection we show how the tadpole diagram in $\Phi^{4}$-theory can
be derived from string theory.
The starting point will be, this time, the color ordered $M$-tachyon planar amplitude
at $h$ loops (\ref{AMh}) specialized  to the case $M=2$ and $h=1$. The
projective invariance can be exploited fixing
\[
V'_{1}(0) = z_{1}=1, \,\,\,\,\,\,\, \eta= 0, \,\,\,\,\,\,\,\,\,
\xi \rightarrow \infty
\]
so that the amplitude becomes
\beqa
A_{2} (p_{1},p_{2}) & =&  N \mbox{Tr} \left[ \lambda^{a_{1}}\lambda^{a_{2}} 
\right] C_{1} \left[ 2 g_{s} ( 2 \alpha' )^{\frac{d-2}{4}} \right]^{2}
\int_{0}^{1} \frac{dk}{k^{2}} \left[ - \frac{1}{2 \pi} \mbox{log} k 
\right]^{-\frac{d}{2}} \prod_{n=1}^{\infty} \left( 1 -k^{n} \right)^{2-d}
\nonumber \\
 & \times& \int_{k}^{1} \frac{dz}{z}  \left[ 
\frac{ \exp{ {\cal G}  ^{(1)} ( 1, z )}}{ \sqrt{z} } 
\right]^{2 \alpha' p_{1} \cdot p_{2} }. 
\eeqa

We have imposed on the
$V_{i}$'s $(i=1,2)$ the condition $V'_{i}(0)=z_{i}$ \cite{DMLRM} and have
renamed $z_{2}=z$.

We would like now to stress that if we want to reproduce diagrams
of scalar field theories we have to ensure that {\it only} tachyon states
propagate in the loops of string amplitudes. In fact this condition is
fulfilled if small values of the multiplier
$k$ are considered: indeed this parameter plays exactly the same role
as the multiplier $A$ in the tree level amplitudes. 
Therefore an expansion in powers of  $k$ is performed
keeping the most divergent terms. In so doing we get
\beq
A_{2}^{(1)} (p_{1},p_{2})= \frac{N}{2} \frac{1}{(4 \pi)^{d/2}} 
\frac{1}{(2 \alpha')^{d/2}}
\left[ 2 g_{s } ( 2 \alpha')^{\frac{d-2}{4}} \right]^{2} 
  \int_{0}^{1} \frac{dk}{k^{2}} \left[ - \frac{1}{2} \log k 
\right]^{-\frac{d}{2}} \int_{k}^{1} \frac{dz}{z} e^{2 \alpha'
{\cal G}^{(1)} ( 1, z )}
\eeq
where the Green function, in the limit we are considering, is
\beq
{\cal G}^{(1)} ( z_{1}, z_{2} ) = \log (z_{1}-z_2) - \frac{1}{2}
\log z_1 z_2 + \frac{ \log^{2} z_1/z_{2} }{2 \log k}
+ O(k)     \label{1g}
\eeq

Our aim is to identify the right limit to get the tadpole diagram in
Fig. \ref{tad}.
 
\begin{figure}[h]
\begin{center}
\parbox{12 cm}{\epsfxsize 12cm \epsfbox{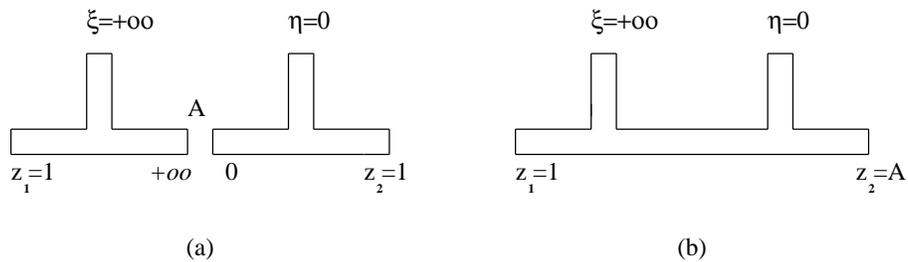}}
\end{center}
\caption{Sewing for the tadpole diagram}
\label{tad1}
\end{figure}

Starting from two three-vertices, we sew the leg $0$ with the 
leg $+\infty$ according to the Fig. (\ref{tad1}).

Such a sewing is performed by considering again the projective transformation
$S(z)=Az$, which has $+\infty$ and $0$ as fixed ponts and which transforms
$z_{2}= 1$ in the second vertex in Fig. (5a) in the multiplier $A$ getting
the configuration shown in Fig. (5b).

The next step consists in performing a limit in which $z_{2} \rightarrow
z_{1}$, i.e. in which $A \rightarrow 1$ with $\alpha' \mbox{log}
(1-A) \rightarrow 0$, as said before. In this limit we should get 
the tadpole diagram. Indeed we have:
\[
A_{2}^{(1)}(p_{1},p_{2}) =  2N \frac{1}{( 4 \pi )^{d/2}} g_{s}^{2} 
(2 \alpha')^{d/2} \int_{0}^{1} \frac{dk}{k^{2}} \left[ -\frac{1}{2} \log
k \right]^{-d/2}  \int_{k}^{1} \frac{dA}{A} e^{ 2 \alpha' p_{1} \cdot p_{2}
[ \log (1-A) - \frac{1}{2} \log A ] }  \nonumber
\] 
\beq
{} =  \frac{2N}{( 4 \pi)^{d/2}} \frac{1}{2 \alpha'}
g^{2}_{s} \int_{0}^{1} \frac{dk}{k^{2}} \left[ -\frac{1}{2} \log k
\right]^{-d/2} \,+O\left( k \right)  \label{limA2}
\eeq
By defining:
\[
x = - \alpha' \log k
\]
with $0 \leq x \leq +\infty $, we can rewrite (\ref{limA2}) as follows:
\beqa
A_{2}^{(1)}(p_{1},p_{2}) & = & \frac{ N}{ ( 4 \pi )^{d/2} } 
\left[ 4 g_{s} ( 2 \alpha'
)^{d/2 - 2} \right] \int_{0}^{\infty}dx  e^{-xm^{2}} x^{-d/2} \nonumber \\
{} & = & \frac{ N}{ ( 4 \pi )^{d/2} } \lambda_{\phi^{4}}
\int_{0}^{\infty}dx  e^{-xm^{2}} x^{-d/2} 
\eeqa

By using the matching condition established at the tree level (\ref{matching})
we get from string theory the tadpole diagram of $\Phi^4$- theory.

\subsection{Candy diagram}

\input epsf

\begin{figure}[h]
\begin{center}
\parbox{6 cm}{\epsfxsize 6cm \epsfbox{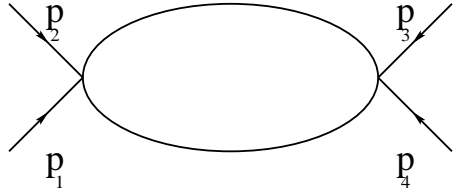}}
\caption{Candy diagram}
\label{candy}
\end{center}
\end{figure}

Let us now derive the {\em candy} diagram from
the four-tachyon one-loop amplitude:
\[
A_{4}^{(1)} (p_{1}, p_{2}, p_{3}, p_{4}) =  \frac{N}{( 4 \pi )^{d/2}} 
\mbox{Tr} \left[ \lambda^{a_{1}}
\lambda^{a_{2}} \lambda^{a_{3}} \lambda^{a_{4}} \right] 
 \frac{1}{( 2 \alpha^{'} )^{d/2} }
\left[ 2 g_{s} ( 2 \alpha^{'} )^{\frac{d-2}{4}} \right]^{4}
\]
\beq
{} \times \int_{0}^{1} \frac{dk}{k^{2}} \left[ -\frac{1}{2} \mbox{log} k
\right]^{-\frac{d}{2}} \int_{k}^{1} \frac{dz_{4}}{z_{4}}
\int_{z_{4}}^{1} \frac{dz_{3}}{z_3} \int_{z_{3}}^{1} 
\frac{dz_{2}}{z_2} \prod_{i<j=1}^{4} \left[ 
\frac{ \exp{ ( {\cal G}(z_{i},z_{j} )) }}{\sqrt{z_i z_j}} 
\right]^{2 \alpha'p_{i} \cdot p_{j}} \label{candy}
\eeq
where we have performed the choice $V'_{i}(0)=z_i$ and, in particular,
we have fixed $V'_{1}(0)=z_1=1$ \cite{DMLRM}.

\begin{figure}[h]
\begin{center}
\parbox{8 cm}{\epsfxsize 8cm \epsfbox{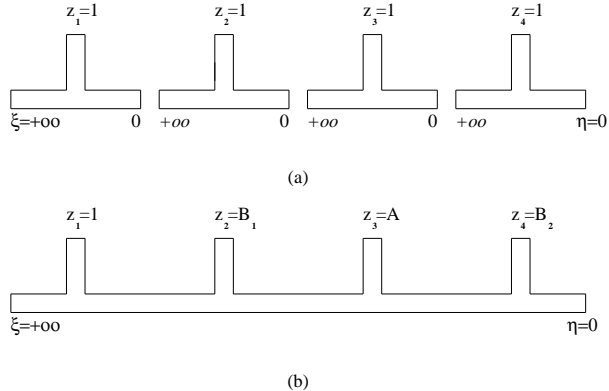}}
\caption{Sewing for the candy diagram}
\label{candy1}
\end{center}
\end{figure}

The diagram relative to this amplitude can be obtained by means of the sewing
procedure illustrated in Fig. \ref{candy1}.

\input epsf

The four-particle vertices of the candy diagram can be generated by
the corner of the moduli space where the
Koba-Nielsen variables $z_{1} \rightarrow z_{2}$ and $z_{3} \rightarrow z_{4}$.
This is performed by considering the limit in which the multipliers 
$B_{i}$ $(i=1,2) \rightarrow 1$. We stress here that, in this limit, the Green
functions ${\cal G} (z_{1},z_{2})$ and ${\cal G} (z_{3},z_{4})$ result to
be divergent and we regularize them by introducing
a cut-off $\epsilon$ on the world-sheet so that $B_{i}=1-\epsilon$. In this
limit the length of the strips connecting the three-vertices become very
short and in this way the four-particle vertices of the diagram 
in consideration are generated.
Furthermore, in order to select in the loop only the lightest states, we also
take the limit in which the multiplier $A \rightarrow 0$, and, after 
having performed both the limits, we send the cut-off to zero 
in all the regular expressions.

From these geometrical considerations that shed light on the different roles
played by the multipliers $A$-like and $B$-like, we select the following
corner of the moduli space reproducing the candy diagram of $\Phi^4$-theory:

\beq
A \rightarrow 0 \hspace{2cm} B_{i}=1-\epsilon \rightarrow 1   \label{corner} 
\label{corncan}
\eeq

Let us now evaluate the amplitude (\ref{candy}) in the corner (\ref{corner}).

The first step consists in rewriting, in this region of the moduli space,
the measure and the integration region in the amplitude (\ref{candy}).

The ordering of the Koba-Nielsen variables determines the integration
regions of the multipliers $A$ and $B_{i}$  in terms of which the whole
amplitude is expressed, after the sewing. More precisely, in the limits
(\ref{corncan}), one gets:

\beqa
&&\int_0^1\frac{d\, k}{k^2}
\int_k^1\frac{d\, z_2}{z_2}\int_k^{z_2}\frac{d\, z_3}{z_3}
\int_k^{z_3}\frac{d\, z_4}{z_4}\simeq 
\int_0^1\frac{d\, k}{k^2}\int_k^1
d B_1\int_{k}^{1}\frac{d\, A}{A}\int_{k}^{1}
d\, B_2\nn\\
&&\simeq\int_0^1\frac{d\, k}{k^2}\int_{k}^1\frac{d A}{A}+
\, O \left(k\right)
\eeqa

For this diagram, the proper times associated to the single propagators
in the loop, are identified with the Schwinger parameters

\beq
t_1 = - \alpha' \log k/A \hspace{2cm} t_2 = -\alpha' \log A 
\eeq
where $k$ has to be understood as the proper time of the whole
loop.

The Green functions defined in (\ref{1g}), in this limit, simplify as
follows:
\beqa
&&2 \alpha' {\cal G} (z_{1},z_{2}) = 2 \alpha' \log \epsilon \,\, , \nn\\
&&2 \alpha' {\cal G} (z_{3},z_{4}) = 2 \alpha' \log \epsilon \,\, , \nn\\
&&2 \alpha' {\cal G} (z_{1},z_{3}) =2 \alpha' {\cal G} (z_{1},z_{4}) = 
2 \alpha' {\cal G} (z_{2},z_{3}) =2 \alpha' {\cal G} (z_{2},z_{4}) \, \, ,\nn\\
&&2 \alpha' {\cal G} (z_{1},z_{3}) = -\alpha' \log A - 
\frac{(-\alpha' \log A)^{2}}{- \alpha' \log k}    \label{Gfun} \,\, .
\eeqa

In particular the Green function $2 \alpha' {\cal G}(z_1, z_3)$, when
written in terms of the Schwinger parameters, becomes
\beq
2 \alpha' {\cal G} (z_1, z_3) = t_2 - \frac{{t_2}^{2}}{t_1 + t_2} \,\, .
\eeq

By expressing the full amplitude in terms of $t_{1}$ and $t_{2}$ one gets:
\beqa
&& A_{4}^{(1)} (p_{1}, p_{2}, p_{3}, p_{4}) = \frac{N}{(4 \pi)^{d/2}}
\frac{1}{2} d^{a_1 a_2 l}d^{a_3 a_4 l} \left[ 2^{6} g^{4}_{s} (2 \alpha')^{d-4} \right] \nonumber \\
&& \int_{0}^{\infty} dt_1 \int_{0}^{\infty} dt_{2} (t_{1} + t_{2} )^{-d/2}
e^{-m^{2} (t_1 + t_2)} e^{-(p_1 + p_2)^2  \left[
 t_2 - \frac{{t_2}^{2}}{t_1 + t_2}\right]}
\eeqa

Once again we have the right result in field theory by using the
matching condition (\ref{matching}).

\section{Two-loop $\Phi^{4}$-diagrams from string amplitudes}
\setcounter{equation}{0}
\subsection{Sunset-diagram}

In this subsection we study the field theory limit of the two-tachyon two-loop
amplitude. It is obtained through  the general expression 
(\ref{AMh}) considered for $M=h=2$: 
 
\beq
A_{2}^{(2)}(p_1,p_2)=N^2 Tr \left[ \lambda^{a_1}\lambda^{a_2}
\right]
C_2 N_0^2
\int \left[dm\right]_2^2 \left[ \frac{ \exp {\cal G}^{(2)}(z_1,\,z_2)}{
\sqrt{ V_1^{'} (0)\, V_2^{'}(0)} } \right]^{2\alpha^{'} p_1\cdot p_2} \,\, ,
\label{2sun}.
\eeq
where the two-loop expressions for $V^{'}_i(0)$ are
given by \cite{DMLMR}:

\beq
(V_{i}^{'}(0) )^{-1}=\left| \frac{1}{z_i - \rho_a}-\frac{1}{z_i - \rho_b}\right|
\label{vi}
\eeq
with  $\rho_a$ and $\rho_b$ depending on the position of $z_i$ and being the 
two fixed points on the left and on the right hand of $z_i$.

For this amplitude we would like to select the regions of moduli space that, in
the field theory limit, reproduce the scalar diagrams of $\Phi^4$ theory, 
in particular in this section we consider the {\em sunset}-diagram, depicted
in Fig. ({\ref{suns}).
\input epsf

\begin{figure}[h]
\begin{center}
\parbox{7 cm}{\epsfxsize 7cm \epsfbox{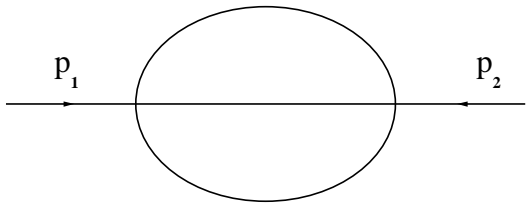}}
\caption{Sunset diagram}
\label{suns}
\end{center}
\end{figure}

Also in this case, the first step consists in making a  
limit that could select the tachyon particles
in the internal string loop. This is achieved considering the amplitude in the 
limit of small multipliers $k_{\mu}$ and keeping the most divergent 
contribution. 

In this limit the various expressions appearing in the 
amplitude take a very simple form. In particular, the Green functions
become \cite{DMLMR}
  
\beqa
&&{\cal G}^{(2)}(z_i,\,z_j)= \log(z_i\, -\, z_j)+\frac{ \log^2 T_{i\,j}\log k_2 +
\log^2 U_{i\,j}\log k_1 - 2 \, \log T_{i\,j} \log U_{i\,j} \log S }{
2(\log k_1 \log k_2 - \log^2S)}\nn\\
&&
\label{gf}
\eeqa

with

\beq
S=\frac{(\eta_1 -\eta_2)(\xi_1-\xi_2)}{(\xi_1-\eta_2)(\eta_1-\xi_2)} \hspace{1cm}
T_{ij}=\frac{(z_j -\eta_1)(z_i-\xi_1)}{(z_j-\xi_1)(z_i-\eta_1)} \hspace{1cm}
U_{ij}= \frac{(z_j -\eta_2)(z_i-\xi_2)}{(z_i-\eta_2)(z_j-\xi_2)} \nn
\eeq

The measure, once used the projective invariance to fix $z_1=1$, 
$\xi_2=+\infty$ and $\eta_2=0$, becomes:

\beq
\left[ d m\right]_{2}^{2}= \frac{dz_2}{ \prod_{i=1}^{2} V_i^{'}(0)}
\prod_{\mu=1}^{2} \frac{d k_{\mu}}{k_{\mu}^{2}} \frac{d \xi_1 d\eta_1}{
(\xi_1 -\eta_1)^2} \left[det \left( -i \tau_{\mu\nu}\right)\right]^{-d/2}  
\label{2ms}
\eeq
where the period matrix $\tau_{\mu\nu}$, in the limit of small multipliers,
is given by:  
   
\beq
det(-i \tau_{\mu\nu})= \frac{1}{4 \pi^2}\left[ \log k_1\, \log k_2\, -\, \log^2 S \right] .
\label{tau}
\eeq

The second step consists in selecting the regions of moduli
that, in the field theory limit, reproduce the sunset-diagram of $\Phi^4$
scalar theory. We cut the 
two-loop string diagram in all possible three-vertices, as shown in Fig. 
(\ref{sunset}). Again, by using the projective invariance, we fix the legs of 
the all three vertices at $+\infty,\,1,\,0$ and we sew them using
suitable projective transformation having as  fixed points  the 
values of the legs we are sewing.

\input epsf

\begin{figure}[h]
\begin{center}
\parbox{8 cm}{\epsfxsize 8cm \epsfbox{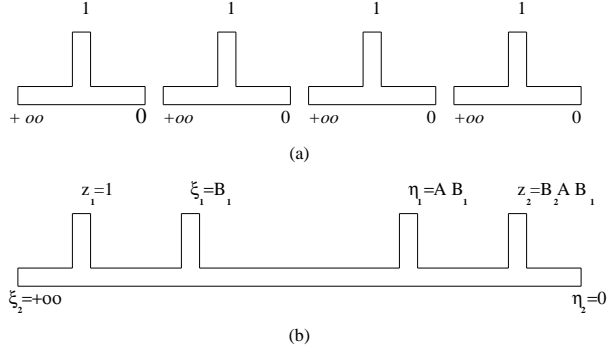}}
\caption{ Sewing for the sunset diagram}
\label{sunset}
\end{center}
\end{figure}

In the case we are here considering we only  need the 
following transformation
\[ 
S(z)=A \, z \hspace{2cm} \mbox{and} \hspace{2cm} S(z)=B_i\, z
\]
with $i=1,2$.  

At the end of this procedure we reach the configuration depicted in Fig. 
(9b), i.e.:

\beqa
&& z_1=1 \hspace{1cm} z_2=A B_1 B_2  \hspace{1cm} \xi_2=+\infty \nn\\
&&\eta_2=0 \hspace{1cm} \xi_1=B_1 \hspace{1cm}  \eta_1=A B_{1}
\label{cornsun}
\eeqa

From (\ref{cornsun}) we  see that the limits to take into account 
for having the quartic vertices of the sunset,
are the ones in which $z_2\rightarrow \eta_1$ and $z_1\rightarrow \xi_1$.
They are simply realized considering a configuration in which 
the $B_i$'s are close to 1, which, in our regularization scheme, 
means to introduce a cut-off on the world sheet, such that $B_i=1-\epsilon$.

Furthermore, in order to select scalars in the other sewn legs, we also 
perform the limit $A\rightarrow 0$. 

In this corner of moduli space we have the following ordering:

\[
\xi_2=+\infty \gg z_1=1 > \xi_1\gg \eta_1 >  z_2 \gg\eta_2=0 .
\] 
and let us now evaluate the amplitude in this corner.

The Green function ${\cal G} (z_1,z_2)$, defined in (\ref{gf}), takes 
the following form:

\beqa
&& 2 \alpha^{'} {\cal G}(z_1\,,\,z_2)\simeq 2\alpha^{'}\log (1-B_1 B_2\, A)\nn\\
&&+ \frac{ \left[ \alpha^{'} \log(\epsilon^2)+\,
\alpha^{'}\log A \right]^2 \log k_2 + 
\left( \alpha^{'} \log A\right)^2 \log k_1- 2 
\left[ \alpha^{'} \log(\epsilon^2) +\alpha^{'}\log A \right] 
\left(\alpha^{'} \log A \right)^2}{ \left( \alpha^{'}\log k_1\right)
\left( \alpha^{'} \log k_2 \right)-\left( \alpha^{'} \log A \right)^2}
\nonumber\\
&&\nn \\
&&\simeq\frac{ \left(\alpha^{'} \log A\right)^2 \left( \alpha^{'} \log k_2
\right) +\left(\alpha^{'} \log A\right)^2\left( \alpha^{'}\log k_1\right) -2
\left(\alpha^{'} \log A\right)^3}{\left(- \alpha^{'}\log k_1\right)
\left(- \alpha^{'}\log k_2\right)-\left( \alpha^{'} \log A\right)^2} \,\,\, ; 
\eeqa
the local coordinates (\ref{vi}) become
\beqa
&&\left(V_1^{'}(0)\right)^{-1}=\left| \frac{1}{z_1-\xi_2}-\frac{1}{z_1-\xi_1}
\right|=\frac{1}{\epsilon}\nn\\ 
&&\left(V_2^{'}(0)\right)^{-1}=\left| \frac{1}{z_2-\eta_2}-\frac{1}{z_2-\eta_1}
\right|=\frac{1}{A \epsilon}\nn\\
\eeqa
while the measure results to be:
\beq
\left[dm\right]_2^2= \prod_{i=1}^2 \left[ d B_i\, \frac{d k_i}{k_i^2}\right]
\frac{dA}{\epsilon^2} \left(4 \, \pi^2\right)^{d/2} 
\left[\log k_1\log k_2-\log^2 A\right]^2
\label{measun}
\eeq

In our field theory limit
we integrate $A$-like variables on the lower boundary of the 
integration region and $B$-like variables in the whole integration region
of the multipliers, i.e. $[0,1]$, as already done in the
tree level case. 

The Schwinger parameters of the field theory diagram are related, 
as explained in the previous sections, to the proper times of 
the propagators used in the sewing. In this case we get:

\beq
t_1=-\alpha^{'}\log A\hspace{2cm} t_i=-\alpha^{'}\log\frac{k_1}{A}
\label{schsun}
\eeq
with $i=2,3$. Expressing the whole amplitudes in terms of these parameters,
we rewrite the exponential involving the Green function as:

\beq
\left[\frac{\exp {\cal G}^{(2)}(z_1\, , z_2)}{\sqrt{V_1^{'}(0) V_2^{'}(0)}} 
\right]^{2\alpha^{'} p_1\cdot p_2}= 
\exp \left\{ - p_1^2 \left[ t_1- \frac{t_1^2 \, (t_2+t_3)}{t_2\,t_3+t_1\,t_2+t_1\,t_3}
\right] \right\}
\label{gftsun}
\eeq
and the measure
\beq
\left[ dm\right ]^2_2=\left(\alpha^{'}\right)^{d/2-3} \prod_{i=1}^{3}d t_i
\prod_{i=1}^2 d \, B_i \left[t_1\,t_2+t_1\,t_3+\,t_2\,t_3\right]^{-d/2}
e^{-m^2\left(t_1+t_2+t_3\right)} e^{-2 \alpha' \log \epsilon}
\label{tmeasun}
\eeq

In the expression of the measure it appears an explicit dependence on
the cut-off $\epsilon$, that disappears in our regularization scheme.
Finally, by inserting these expressions in the amplitude (\ref{2sun})
and writing explicitly the values of the normalization constants
we get:
\beqa
 A_{2}^{(2)}(p_1\,,p_2)& =&  \frac{N^2}{\left(4\pi\right)^d}
\left[ 4 g_s^2 \left(2 \alpha^{'} \right)^{d/2-2}\right]^2
\int_{0}^{+\infty}\prod_{i=1}^3 d\, t_i
\left[ t_1\,t_2+t_1\,t_3+\,t_2\,t_3\right]^{-d/2}\nn\\
 &\times& \exp \left\{ - p_1^2 \left[  
\frac{t_1\, t_2\,t_3 }{ t_2\,t_3+t_1\,t_2+t_1\,t_3} \right] \right\}
\label{sunset}
\eeqa

This expression, once the matching condition (\ref{matching}) is 
used, is coincident, including the overall factor, with the corresponding one 
obtained in field theory starting from the action (\ref{phi4}).

\subsection{Double candy diagram}
\input epsf
\begin{figure}[h]
\begin{center}
\parbox{7 cm}{\epsfxsize 7cm \epsfbox{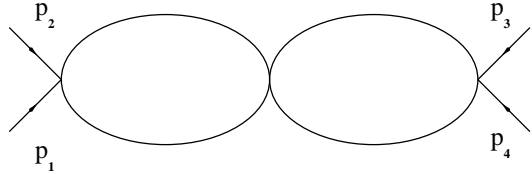}}
\end{center}
\label{2candy}
\caption{ Double-candy}
\end{figure}
In this subsection we show how to get the {\em double-candy} diagram  of 
$\Phi^4$ theory, Fig. (10}), starting from the 
two-loop four-tachyon amplitude in bosonic string theory.

The starting point is again the master formula  
 (\ref{AMh}) with $M=4$ and $h=2$:
 
\beq
A_{4}^{(2)}(p_1 p_2 p_3 p_4)=N^2 Tr \left[ \lambda^{a_1} \lambda^{a_2}
\lambda^{a_{3}} \lambda^{a_4}
\right]
C_2 N_0^4
\int \left[dm\right]_2^4 \prod_{i<j}\left[ \frac{ \exp {\cal G}^{(2)}(z_i,\,z_j)}{
\sqrt{ V_i^{`} (0)\, V_j^{`}(0)} } \right]^{2\alpha^{'} p_i\cdot p_j}
\label{2ta}
\eeq
where the expressions for $V^{'}_i(0)$ are given by (\ref{vi}). 

As in the case of the sunset-diagram, we expand the previous amplitude  
for small values of the  multipliers $k_{\mu}$ keeping
the most divergent contribution that is the one corresponding to
the tachyon state and again the Green functions reduce to the
form given in (\ref{gf}).

The measure, once used the projective invariance to fix $z_4=1$, 
$\xi_2=+\infty$ and $\eta_2=0$, becomes:

\beq
\left[ d m\right]_{2}^{4}=\prod_{i=1}^{3}
\frac{d z_i}{V_i^{'}(0)}
\prod_{\mu=1}^{2} \frac{d k_{\mu}}{k_{\mu}^{2}} \frac{d \xi_1 d\eta_1}{
(\xi_1 -\eta_1)^2} \left[det \left( -i \tau_{\mu\nu}\right)\right]^{-d/2}  
\label{2ms}
\eeq
where the period matrix $\tau_{\mu\nu}$ in the limit of small multipliers
is given by Eq. (\ref{tau}).

Let us now identify the corner of the moduli space that, in the field theory 
limit, reproduces the two-loop candy diagram, according to our procedure.
In Fig. (11) it is shown the final configuration that we reach 
applying the sewing procedure with the following projective 
transformations:    

\beq
\begin{array}{ccc}
S_i= B_i \, z &  \hat{S}_1=A_1\, z  & \hat{S}_2=A_2\, z
\end{array}
\label{prt}
\eeq
with $i = 1,2,3$.

Once the sewing procedure is completed, the Koba-Nielsen variables 
and the moduli of the surface, are expressed in terms of the multiplier of the
transformations. For the sewing configuration shown in Fig. (\ref{2loop1})   
we get the following correspondence 
among multipliers, Koba-Nielsen variables and moduli:

\beq
\begin{array}{cccc}
\xi_2 = \infty & \hspace*{.5cm} \eta_2 =0 & \hspace*{.5cm} \xi_1=B_1 B_2 A_1
&\hspace*{.5cm} \eta_1= (1-A_2) A_1 B_1 \\
              &                         &               &\\
z_1= A_1 B_1 & \hspace*{.5cm} z_2=\left[1-A_2(1-B_3)\right]A_1 B_1 
& \hspace*{.5cm}z_4=1
& \hspace*{.5cm} z_3=B_1
\end{array}             
\label{cor1}
\eeq

Again, if we want to obtain 
the four-particle vertices peculiar of the two-loop candy diagram,
we have to take in consideration the corner of the Koba-Nielsen 
variables characterized by $z_1\rightarrow z_2$, 
$z_3\rightarrow z_4$ and by the modulo
$\xi_1\rightarrow 1$. This configuration is achieved considering 
the limits in which 
$B_i\rightarrow 1$ and introducing the suitable regularizators, when necessary.

Furthermore, considering also the limit $A_i\rightarrow 0$, we select scalar
particles in the other internal legs.

The corner of  moduli space reproducing the $\Phi^4$ scalar diagram 
illustrated in Fig. \ref{2loop1} is

\beq
\begin{array}{lr}
A_{i}\rightarrow 0  & \hspace*{1cm}B_{i}= 1 -\epsilon  \,\,\, .  
\end{array}
\label{cor}
\eeq

Let us now evaluate  the amplitude (\ref{2ta}) in this corner.

The Green functions (\ref{gf}) in the limit (\ref{cor})
take the simple form:

\beqa
&&\alpha^{'} {\cal G} \left(z_1,\,z_3\right)=\alpha^{'} {\cal G} 
\left(z_1,\,z_4\right)=\alpha^{'} {\cal G} \left(z_2,\,z_3\right)=
\alpha^{'} {\cal G} \left(z_2,\,z_4\right) \,\, , \nn\\
&&
\alpha^{'}{\cal G} (z_1,z_3) = -\frac{1}{2} 
\frac{(-\alpha^{'}\log A_2)^2}{-\alpha^{'}\log k_1}-\frac{1}{2}
\frac{(-\alpha^{'} \log A_1)^2}{-\alpha^{'}\log k_2} \, \, , \nn\\
&& \alpha^{'} {\cal G}\left(z_1,\,z_2\right) 
\simeq \alpha^{'}\log\left( A_1 A_2 \epsilon \right) \, \, ,\nn \\
&& \alpha^{'}{\cal G}\left(z_3\, , z_4\right)\simeq \alpha^{'} 
\log( \epsilon)\simeq 0  \,\, .
\label{gco2}
\eeqa

A similar manipulation can be done for the local coordinates $V_i^{'}(0)$
that in the limit (\ref{cor}) become:

\beqa
&&\left(V_1^{'}(0)\right)^{-1}=\left| \frac{1}{z_1 -\eta_1}-\frac{1}{z_1-\xi_1}
\right|
\simeq\frac{1-\frac{A_2}{\epsilon}}{A_1\,A_2}\simeq \frac{1}{A_1\,A_2}\nn\\
&&\left(V_2^{'}(0)\right)^{-1}=\left|\frac{1}{z_2-\eta_1}-\frac{1}{z_2-\xi_2}
\right|\simeq \frac{1}{A_1 \,A_2}\nn  \\
&&\left(V_3^{'}(0)\right)^{-1}= \left|\frac{1}{z_3-\xi_1} -
\frac{1}{z_3-\xi_2} \right|=1
\hspace{1cm}\left(V_4^{'}(0)\right)^{-1}=\left|\frac{1}{z_4-\xi_1} -
\frac{1}{z_4-\xi_2} \right|=1\nn\\
&&
\label{vic}
\eeqa
and for the measure:

\beq
\left[ d m\right]^4_2 =  \left[ \prod_{i=1}^{2} \frac{d A_i}{A_i}\right]
\prod_{i=1}^{3} d B_i \prod_{i=1}^{2} 
\frac{d k_i}{k_i^2}\left( \log k_1\, \log k_2 \right)^{-d/2} \left[ 
\frac{1}{\epsilon^2}\right]\,\, .
\label{mea}
\eeq

As regards the integration region, we observe that the sewing procedure 
determines an ordering of the Koba-Nielsen variables and of the fixed points.
In the case in consideration here we get in fact the following ordering:

\[
\xi_2=+\infty \gg z_4=1 \geq z_3=B_1 \geq z_1=A_1B_1 \gg \xi_1=B_2\,A_1B_1\geq 
\eta_2=0  
\]
and
\[
z_1\geq z_2=\left[ 1-A_2(1-B_3)\right]A_1B_1 > \eta_1=(1-A_2)A_1B_1\,\, .
\]

In the field theory limit (\ref{cor}) we integrate the multipliers 
$B$-like between $0$ and $1$ and the multipliers $A$-like, between $0$ and
$\delta$ being $\delta$, a positive infinitesimal quantity.

The Schwinger parameters in this case 
are related to the $A_i$'s 
by the following 
relations \cite{DMLMR}:

\beq
t_{i+2}=-\alpha^{`} \log A_i \hspace*{1cm}
t_1=-\alpha^{'} \log \frac{k_1}{A_2} \hspace{1cm}
t_2= -\alpha^{'} \log \frac{k_2}{A_1} 
\label{sch}
\eeq
with $i=1,2$.

Rewriting the Green functions in terms of the Schwinger parameters:

\beq
\prod_{ i<j=1 }^{4} \left[
\frac{\exp { {\cal G}^{(2)}(z_i\, , z_j)}}{\sqrt{V_i^{'}(0) V_j^{'}(0)}} 
\right]^{2\alpha^{'} p_i\cdot p_j}= 
\exp \left\{- (p_1+p_2)^2 \left[ \frac{t_1 \, t_4}{t_1+t_4} + 
\frac{t_2 \, t_3}{t_2+t_3} \right] \right\}
\label{gft}
\eeq
and the measure

\beq
\int \left[ d m\right]^4_2= \left(\alpha^{'}\right)^{d-4} \, (2\pi)^d 
\int_{0}^{+\infty}\prod_{i=1}^{4} d t_i
\int_{0}^{1}\prod_{i=1}^3 d B_i e^{-m^2(t_1+t_2+t_3+t_4)} 
 (t_2+t_3)^{-d/2} \,(t_4+t_1)^{-d/2} \,\, . 
\label{met}
\eeq
we get

\beqa
&&A_{4}^{(2)}(p_1\cdots p_4)=\frac{N^2}{(4\pi)^2} d^{a_1\,a_2\,l} \,
d^{a_3\,a_4\,l}
\frac{ \left[ 2^4 g_s^2 (2\alpha^{'})^{\frac{d-4}{2}}\right]^3}{2^5}
\nn\\
&&\times\int_{0}^{\infty} \prod_{i=1}^4 d t_i e^{  -m^2(t_1+t_2+t_3+t_4)}
(t_1+t_4)^{-d/2} (t_2+t_3)^{-d/2} e^{-(p_1+p_2)^2 \left[ 
\frac{t_1\, t_4}{t_1\, +\, t_4} + \frac{t_2 \, t_3}{t_2 + t_3}\right]}
\label{am}
\eeqa
where a sum over inequivalent permutations of the external particles 
has been done analogously as in the one-loop candy-diagram case.

Now using the matching condition (\ref{matching}),
we get the same result, including the overall factor, as the one 
obtained in field theory. 

In conclusion, we have used the {\em sewing and cutting} procedure in order to show
how $\Phi^4$-theory diagrams can be reproduced from string amplitudes,
up to two loop-order. The whole information so obtained can be in principle
extendible to Yang-Mills diagrams involving quartic interactions.

\input epsf

\begin{figure}[h]
\begin{center}
\parbox{10 cm}{\epsfxsize 10cm \epsfbox{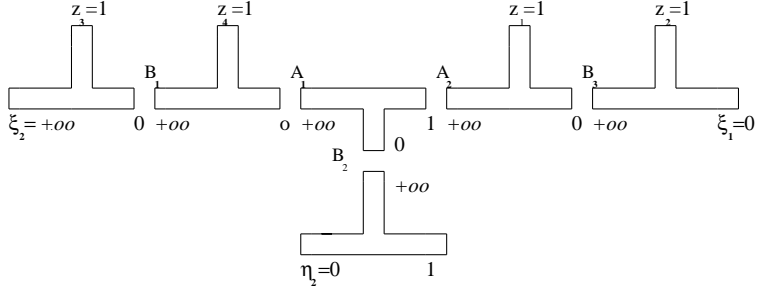}}
\end{center}
\label{2loop}
\caption{Sewing of the two-loop candy diagram}
\end{figure}

\input epsf

\begin{figure}[h]
\begin{center}
\parbox{11 cm}{\epsfxsize 11cm \epsfbox{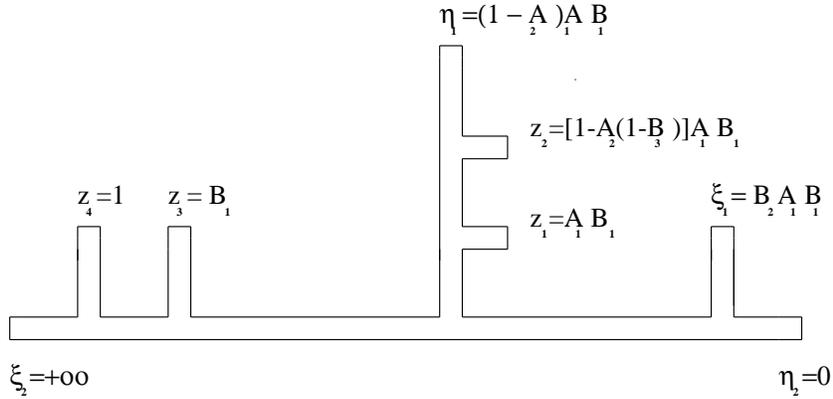}}
\end{center}
\label{2loop1}
\caption{Sewing configuration for the two-loop candy diagram}
\end{figure}

\vspace{2.5cm}

{\bf Acknowledgement}

We would like to thank P. Di Vecchia and M.G. Schmidt for helpful discussions.
Furthermore, we particularly thank A. Frizzo, L. Magnea and R. Russo, with whom
we have continuously exchanged information about our and their techniques
for getting scalar field theory amplitudes. They are publishing their
results in a forthcoming work. 

R.M. and F.P. thanks respectively Universit\`a di Napoli and NORDITA 
for their kind hospitality during different stages of this work.

\end{document}